\documentclass[submission,copyright,creativecommons]{eptcs}
\providecommand{\event}{ICLP/LPNMR 2024} 
\usepackage{underscore}           

\usepackage{xspace}
\usepackage{amsmath}
\usepackage{xcolor}
\usepackage{graphicx}   

\newcommand{\clingo}{{\sc clingo}\xspace}
\newcommand{\dlv}{{\sc dlv}\xspace}
\newcommand{\anthem}{{\sc anthem}\xspace}
\newcommand{\vampire}{{\sc vampire}\xspace}
\newcommand{\selp}{{\sc selp}\xspace}
\newcommand{\tabeql}{{\sc tabeql}\xspace}
\newcommand{\aspide}{{\sc aspide}\xspace}
\newcommand{\ag}{{\sc abstract gringo}\xspace}
\newcommand{\mg}{{\sc mini-gringo}\xspace}
\newcommand{\ruleo}{\;\hbox{:-}\;}
\newcommand{\boldp}{\mathbf{p}}
\newcommand{\rar}{\rightarrow}

\begin{document}

\title{Relating Answer Set Programming and\\Many-sorted Logics for Formal Verification}

\author{Zachary Hansen
\institute{University of Nebraska Omaha\\Omaha, Nebraska\\929 S. 70th Plz \#25, Omaha NE, 68106}
\email{zachhansen@unomaha.edu}
}

\def\titlerunning{Relating ASP and Many-sorted Logics}
\def\authorrunning{Hansen}

\maketitle










\section{Introduction}
Answer Set Programming (ASP) is an important logic programming paradigm within the field of Knowledge Representation and Reasoning.
As a concise, human-readable, declarative language, ASP is an excellent tool for developing trustworthy (especially, artificially intelligent) software systems.
However, formally verifying ASP programs offers some unique challenges, such as
\begin{enumerate}
    \item a lack of modularity (the meanings of rules are difficult to define in isolation from the enclosing program),
    \item the ground-and-solve semantics (the meanings of rules are dependent on the input data with which the program is grounded), and
    \item limitations of existing tools.
\end{enumerate}
My research agenda has been focused on addressing these three issues with the intention of making ASP verification an accessible, routine task that is regularly performed alongside program development.
In this vein, I have investigated alternative semantics for ASP based on translations into the logic of here-and-there and many-sorted first-order logic.
These semantics promote a modular understanding of logic programs, bypass grounding, and enable us to use automated theorem provers to automatically verify properties of programs.
%


\section{Background}
The stable model semantics of logic programs can be expressed in a variety of ways, each of which offers unique insights and utility~\cite{lif10}.
For instance, logic programs can sometimes be viewed as ``shorthand" for propositional, first-order, default, or autoepistemic theories.
These translational approaches (which typically involve a syntactic transformation from ASP rules into a ``formula representation" that uses the syntax of first-order logic) offer some advantages over their fixpoint relatives.
In particular, semantics that let us bypass the issue of grounding are very convenient for program verification purposes.
They allow us to disregard the specifics of the input data with which a program is paired when assessing the independent meaning of the program.

Clark's Completion and its subsequent extensions~\cite{Clark1978,liflusch} provide a useful translational semantics for a broad class of programs satisfying the \emph{tightness} condition\footnote{A tight program has a predicate dependency graph without positive cycles~\cite{falilusc20a,fages94a}.}.
Under these semantics, the logic program
$$
p \ruleo q. \quad p \ruleo r.
$$
is understood as the first-order theory
$$
p \leftrightarrow q \vee r
$$
which reflects the minimality of circumscription and the closely related principle of rational belief (we believe $p$ only if we have to, that is, if and only if $q$ or $r$ holds) found in the autoepistemic intuitions behind answer set programming~\cite{gelkahl14}.
However, completion semantics are only applicable to programs of a limited form -- first-order logic cannot, for example, correctly capture the transitive closure of a binary relation without non-standard assumptions like the Closed World Assumption.
%
This restriction can potentially be circumvented or at least relaxed through innovation in the areas of local tightness~\cite{fanliftem24}, loop formulas~\cite{leelif03}, tightening~\cite{wall93}, and ordered completion~\cite{asuchezhazho15}.

More recently, a surprisingly deep connection between the logic of here-and-there~\cite{heyting30a} and ASP has yielded interesting results.
The logic of here-and-there extends intuitionistic logic with the axiom schema 
$$F \vee (F \rightarrow G) \vee \neg G$$ 
of which the most commonly employed consequence is the weak law of excluded middle:
\begin{gather}
    \label{eq:weak.ex.middle}
    \neg F \vee \neg\neg F.
\end{gather}

Here-and-there has a number of useful characteristics that make it applicable to the study of ASP~\cite{pearce06}.
First, it acts as a well-behaved monotonic basis for equilibrium logic, which provides a non-monotonic inference relation that captures and generalizes the semantics of a broad class of ASP programs to full propositional logic (recall that while ASP rules have a very limited syntactic form, propositional formulas can be very complex).
Furthermore, intuitionistic logic cannot be strengthened more than here-and-there and still be strictly contained in classical logic, making it the strongest, well-behaved classical logic available.
Finally, the extension to ASP programs with variables is naturally covered by quantified formulas interpreted under the semantics of here-and-there.

One of the most fruitful consequences of this connection is the study of \emph{strong equivalence}~\cite{LifschitzPearceValverde01}.
The condition of strong equivalence states that two programs, $\Pi_1$ and $\Pi_2$, are strongly equivalent if $\Pi_1 \cup \Pi$ and $\Pi_2 \cup \Pi$ have the same answer sets for any program $\Pi$.
This condition is useful because it tells us that our two programs are truly interchangeable: no matter which program they form a subcomponent of, they can be swapped out without affecting the enclosing program's answer sets.
It has been shown that the strong equivalence of two programs can be established by deriving the formula representation of each program from the formula representation of the other within the logic of here-and-there; for propositional programs this can be done in exponential time~\cite{LifschitzPearceValverde01}.
However, more complex ASP languages require more sophisticated translations into formula representations, as well as extended deductive systems.
For the remainder of this paper, we will focus on a theoretical ASP language known as \mg.

The answer set solver \clingo implements the language \ag~\cite{ag15}, whose semantics are defined by a translation ($\tau$) into the infinitary propositional logic developed by Truszczy\'{n}ski~\cite{tru12}.
\mg is an expressive fragment of \ag which supports negation, arithmetic, intervals, and basic choice rules~\cite{fanliftem24}.
The semantics of the language can be captured by a syntactic transformation ($\tau^*$) into first-order formula representations, which are interpreted under the HTA (here-and-there with arithmetic) deductive system~\cite{fanlif23}.
An equivalent (and arguably simpler) characterization of these semantics are defined by the SM operator~\cite{feleli11a}. 
%
This operator transforms a program's ($\Pi$) formula representation ($\tau^*\Pi$) into a second-order sentence ($SM_\boldp[\tau^*(\Pi)]$) in which predicate quantification is used to minimize belief in the list $\boldp$ of {\em intensional} predicates (similar to circumscription). 
When $\boldp$ contains all the predicates occurring in the logic program, then the models of $SM_\boldp[\tau^*(\Pi)]$ correspond to the answer sets of $\Pi$.
Bartholomew and Lee extend the concept of the $SM$ operator to intensional functions~\cite{bartlee19}.
This is once again rooted in the logic of here-and-there, where interpretations can be viewed as pairs (of ``worlds") $\langle H, T \rangle$.
Predicate minimization through the $SM$ operator is achieved by mandating that $p^H \subseteq p^T$ for every $p \in \boldp$, similarly, for each intensional function $f$ we require that $f^H \neq f^T$.

The study of the \mg language is motivated in large part by a desire to use automated reasoning tools for proving the correctness of ASP programs.
\anthem is a software system that converts \mg programs into first-order formulas of two sorts (a supersort which consists of all program terms, and a subsort corresponding to integers) via the $\tau^*$ translation~\cite{falilusc20a}.
It then uses the first-order theorem prover \vampire~\cite{kovvor13a} to check the equivalence of the completion of the program ($COMP[\tau^*(\Pi)]$) to a set of first-order formulas acting as a specification ($S$), under a set of assumptions characterizing the intended program inputs ($A$).
That is, it attempts to establish the universal validity of
$$
A \rightarrow (COMP[\tau^*(\Pi)] \leftrightarrow S)
$$
which, if it succeeds, is proof that the program $\Pi$ implements the specification.
This procedure applies to \emph{io-programs}, that is, ASP programs paired with placeholders as well as input and output predicate symbols.
Any predicate symbols that are neither input nor output symbols are private symbols -- these auxiliary symbols are not essential for understanding the program's ``external" behavior as characterized by the output symbols.
For example, if we take $prime/1$ to be an output symbol, the programs
\begin{align*}
    composite(I*J) &\ruleo I = 2..n, ~J = 2..n.\\
    prime(I) &\ruleo I = 2..n, ~not ~composite(I).
\end{align*}
and
\begin{align*}
    comp(X) &\ruleo X = I*J, ~I = 2..n, ~J = 2..n.\\
    prime(I) &\ruleo I = 2..n, ~not ~comp(I).
\end{align*}
clearly have the same external behavior (the extent of the $prime/1$ predicate) under the assumption that the placeholder $n$ is an integer, despite minor differences in how the auxiliary predicates are defined.
This procedure is only possible for programs satisfying the restrictions of tightness and absence of private recursion.\footnote{A program is tight if its predicate dependency graph has no cycles consisting of positive edges such as $(p, p)$. A program has private recursion if the subgraph induced by private symbols has no cycles.}


\section{Goals}
My interest in the relationship of ASP to many-sorted first-order logic (and, more broadly, the logic of here-and-there) is motivated by its application to software verification.
My long-term research agenda is to help develop an accessible, tool-assisted methodology for formally verifying the correctness of ASP programs.
Clearly, such an agenda would extend beyond the duration of a single dissertation.
For this reason, I've organized the remainder of this research summary into sections showcasing which pieces of this plan have already been addressed (Current Status and Preliminary Results), which pieces I plan to investigate during the remainder of my doctoral program (Ongoing Directions and Expected Achievements), and which pieces I hope to build on top of my eventual dissertation (Conclusions and Future Directions).


\section{Current Status and Preliminary Results}
Since joining the University of Nebraska Omaha (UNO) in Fall 2020, I have been engaged in several research projects under the umbrella topic of \emph{formal verification of ASP programs}.
Within this topic, I have been fortunate to collaborate with researchers at UNO, University of Texas at Austin, and the University of Potsdam.
My frequent collaborators include Dr. Yuliya Lierler (my advisor), Dr. Jorge Fandinno, Dr. Vladimir Lifschitz, Dr. Torsten Schaub, and Tobias Stolzmann (another PhD student writing his dissertation on topics related to \anthem).
Where appropriate, I will distinguish between collaborative work and work I've done independently.

\subsection{Results Presented at Previous Doctoral Consortiums}

\paragraph{Conditional literal semantics}
Conditional literals are a useful feature supported by \clingo.
The following rule (from a Graph Coloring encoding) succinctly expresses that a vertex cannot have no colors assigned to it.
\begin{gather}
\hbox{\tt :- not asg(V, C) : col(C); vtx(V).} \label{eq:coloring.cl}
\end{gather}
In the spirit of Fandinno et al. (2020), Dr. Lierler and I developed a translation from a simple ASP language that is (mostly) a subset of \mg extended with conditional literals to unsorted first-order logic~\cite{hanlie22}.
This provided formal support for the intuition that conditional literals in this language represent nested implications within rule bodies.
For example, the previous rule can be understood as the first-order sentence
\begin{gather}
    \label{eq:coloring.cl.fo}
    \forall V \big(\left(\forall C (col(C) \rar \neg asg(V,C)) \wedge vtx(V)\right) \rar \bot\big).
\end{gather}
We demonstrated that these semantics capture the behavior of \clingo, and used them to prove the correctness of a \emph{k-coloring program}.

\paragraph{Aggregate semantics}
Dr. Lierler, Dr. Fandinno, and I proposed a characterization of aggregate semantics that bypasses the need for grounding~\cite{fan22}. 
Instead, we apply a many-sorted generalization of the SM operator to a set of many-sorted first-order formulas ($\kappa \Pi$) representing a logic program ($\Pi$). 
Aggregates are defined as functions on sets of tuples, whose members are restricted to those tuples satisfying the list of conditions present in the associated aggregate. 
We designed a set of second-order (first-order in the presence of finite aggregates) axioms to define the behavior of sets and aggregate function symbols.
For a class of standard interpretations satisfying assumptions such as a standard interpretation of addition, models of $SM[\kappa \Pi]$ satisfying these aggregate axioms are in one-to-one correspondence with the stable models of $\Pi$.
When $\Pi$ is tight, the second-order characterization ($SM[\kappa \Pi]$) can be replaced by completion ($COMP[\kappa \Pi]$).
Thus, for tight programs with finite aggregates, our proposed semantics define, via first-order logic, the behavior of \clingo aggregates.

\paragraph{Modular proofs of correctness with aggregate constraints}
A key challenge in verifying logic programs is proving the correctness of groups of rules in isolation from the rest of the program.
A ``divide-and-conquer" methodology is very natural  for verification, but applying it to logic programs requires a careful methodology, such as the one proposed by Cabalar, Fandinno, and Lierler (2020). 
They divide their example Hamiltonian Cycle program into various independent modules, whose behavior is captured via the SM operator~\cite{cab20b}. 
We extend their example to the Traveling Salesman problem with the addition of an aggregate constraint~\eqref{rule:tsp} on the cumulative weight of the selected cycle, and use our many-sorted semantics for aggregates to verify the behavior of this constraint independently of the Hamiltonian Cycle program~\cite{fanhanlie22arg}. 
Additionally, we prove the correctness of a Graph Coloring encoding containing choice rules with cardinality bounds.
This project showcases how the modular proof methodology can be extended with our proposed aggregate semantics to argue the correctness of a broad class of logic programs.

\begin{gather}
    \hbox{\tt :- \#sum\{ K,X,Y : in(X,Y), cost(K,X,Y) \} > J, maxCost(J).}
    \label{rule:tsp}
\end{gather}

\subsection{New Results}

\paragraph{Extending \mg with conditional literals}
The \ag fragment Dr. Lierler and I investigated in 2022 was unsorted, forbade double negations, and lacked support for arithmetic operations and intervals.
These are important features supported by the \mg language on which \anthem is based.
We are working on extending the full \mg language presented in Fandinno, Lifschitz, and Temple (2024) with conditional literals.
The translation of conditional literals is largely the same, but proving the correctness of the translation for this extended language is considerably more complicated.
However, doing so allows us to check the strong equivalence of \mg programs with conditional literals.
Furthermore, this work is useful because it acts as a roadmap for the more challenging task of extending \mg and \anthem with aggregates using our many-sorted first-order characterization.
%
%

This research has also yielded some interesting insights into the way conditional literals can be used to eliminate auxiliary predicates.
Doing so makes modular programming, and constructing arguments of correctness about modular programs, considerably easier.
For example, consider a traditional encoding of the Graph Coloring problem without conditional literals:
\begin{align}
    &\hbox{\tt \{asg(V, C)\} :- vtx(V), col(C).} \\
    &\hbox{\tt  :- asg(V, C1), asg(V, C2), C1 != C2.}       \\
    &\hbox{\tt colored(V) :- asg(V, C).} \label{eq:coloring.colored}\\
    &\hbox{\tt :- vtx(V), not colored(V).}   \label{eq:coloring.must.color}    \\
    &\hbox{\tt :- asg(V1, C), asg(V2, C), edge(V1, V2).} 
\end{align}
We can eliminate the auxiliary predicate $colored/1$ by replacing rules (\ref{eq:coloring.colored}-~\ref{eq:coloring.must.color}) with the conditional literal constraint~\eqref{eq:coloring.cl}.
This example hints at a more general property.
Note that the rule~\eqref{eq:coloring.colored} expresses that a property ($colored$) holds for a vertex ($V$) if and only if there exists an element ($C$) such that $V$ is mapped to $C$ by the $asg/2$ predicate.
This is an indirect way of expressing an existential quantification -- in conjunction with rule~\eqref{eq:coloring.must.color}, it expresses that every vertex must be mapped to a color.
This condition can be more concisely represented via conditional literal~\eqref{eq:coloring.cl.fo}, which is classically equivalent to
\begin{align*}
    \forall V \big( vtx(V) \to \exists C (col(C) \wedge asg(V,C)) \big). 
\end{align*}


\paragraph{Axiomatizing new aggregates} 
The many-sorted first-order logic characterization of aggregates Dr. Lierler, Dr. Fandinno, and I developed captured the behavior of \clingo's \texttt{count} and \texttt{sum} aggregates~\cite{fan22}.
Furthermore, for programs adhering to the ASP-Core-2 standard~\cite{aspcore20}, this characterization also captures the ASP-Core-2 semantics.
I have since extended this characterization with \texttt{min}, \texttt{max}, and \texttt{sum+} aggregates.
This required developing first and second-order axiomatizations for the new aggregates and proving that they correctly captured the behavior of \clingo.
These results, alongside significant extensions such as a detailed section integrating our results with Clark's Completion, were recently published in the Journal of Artificial Intelligence Research~\cite{fanhanlie24}.

\paragraph{Recursive aggregates}
The aggregate axiomatization projects described thus far have included a restriction on positive recursion through aggregates.
While such recursion is a comparatively rare scenario in practice, it is very important to the study of strong equivalence.
Since two programs $\Pi_1$ and $\Pi_2$ must have the same answer sets when combined with any program $\Pi$ to be strongly equivalent, any discussion of strongly equivalent logic programs must accommodate the case when $\Pi$ introduces recursion through the aggregates of $\Pi_1$ or $\Pi_2$.
To address this, Dr. Fandinno and I extended our aggregate semantics with intensional function symbols~\cite{fanhan23}.
We treated aggregates as functions on sets of tuples -- these so-called ``set symbols," which map ground terms to sets of tuples of ground terms, were treated intensionally.
We found that our proposed semantics coincides with the semantics of \clingo, but naturally diverged from the semantics of \dlv in the presence of recursive aggregates.
I presented these results at ASPOCP 2023.

\paragraph{Program to program verification}
The original version of \anthem required a specification written in first-order logic.
However, it has since become clear that many ASP programmers would rather write their specifications in the form of a simple, easy-to-read ASP program.
To support this, I developed the AP2P\footnote{\url{https://ap2p.unomaha.edu/}} system on top of the original \anthem.
The theoretical results supporting this procedure were published in TPLP~\cite{fanhanliftem23}.
AP2P allows users to automatically confirm the (external) equivalence of two ASP programs.
This is primarily useful in the refactoring process, wherein a simple program may be successively replaced by more complex programs in the interest of improving performance.
Our system checks that the essential behavior of the program has not been changed during refactoring.

\paragraph{Anthem 2} 
The long-term viability of the prototypical \anthem system was threatened by technical debt such as a lack of documentation, heavy dependence on deprecated Rust nightly features, and a divergence of the system’s behavior from the supporting theory. 
I have been working in collaboration with Dr. Lifschitz and Tobias Stolzmann on a complete overhaul and re-implementation of this prototype that corrects and extends it in several ways. 
First, I have restructured and generalized the verification process to enable a symmetric treatment of program-to-program and program-to-specification verification. 
This positions \anthem as a tool to support refactoring ASP code in addition to its original functionality.
Second, Tobias Stolzmann and I have corrected the internal representation of many-sorted first-order theories and the partial axiomatizations of standard interpretations. 
This eases (in particular) the handling of placeholders in io-programs. 
Third, I have designed and implemented a suite of transformations equivalent in the logic of here-and-there to simplify the formulas being passed to the backend theorem prover. 
Preliminary experiments show substantial improvements in runtime for certain programs.
Fourth, Dr. Lifschitz and I have designed and implemented an improved control language for writing proof outlines, which offers users a considerably more fine-grained control over the formulation of verification tasks. 
%
This moves \anthem away from a one-shot system towards an interactive proof assistant, which is crucial for verifying non-trivial problems.
Finally, I wrote a reference manual to resolve the issue of missing documentation\footnote{The new system and user manual can be found here: \url{https://github.com/potassco/anthem}}. 
%


\section{Ongoing Directions and Expected Achievements}

\paragraph{Recursive aggregates}
Dr. Fandinno and I are currently developing recursive aggregate semantics for \dlv analogous to those we created for \clingo.
We have designed a new translation to a many-sorted first-order language for \dlv programs, that treats default negation in a different manner.
This work suggests that the difference between the treatment of recursive aggregates in \clingo versus \dlv stems from an underlying difference in their treatment of default negation.
These new semantics allow us to define strong equivalence not just between a pair of \clingo programs or a pair of \dlv programs, but between a \clingo program and a \dlv program.
These are exciting results that provide new insights into the differences between these two major solvers.

\paragraph{Extending \anthem with conditional literals}
More theoretical work needs to be done before I can implement a translation for conditional literals within \anthem.
First and foremost, the current results need to be extended to io-programs.
Additionally, while checking strong equivalence does not require tightness, checking external equivalence does.
It is not yet clear to me how the notion of a predicate dependency graph changes in the presence of conditional literals, although first-order dependency graphs are a promising direction to explore~\cite{leemeng11}.

\begin{figure}
    \centering
    \includegraphics[width=0.35\linewidth]{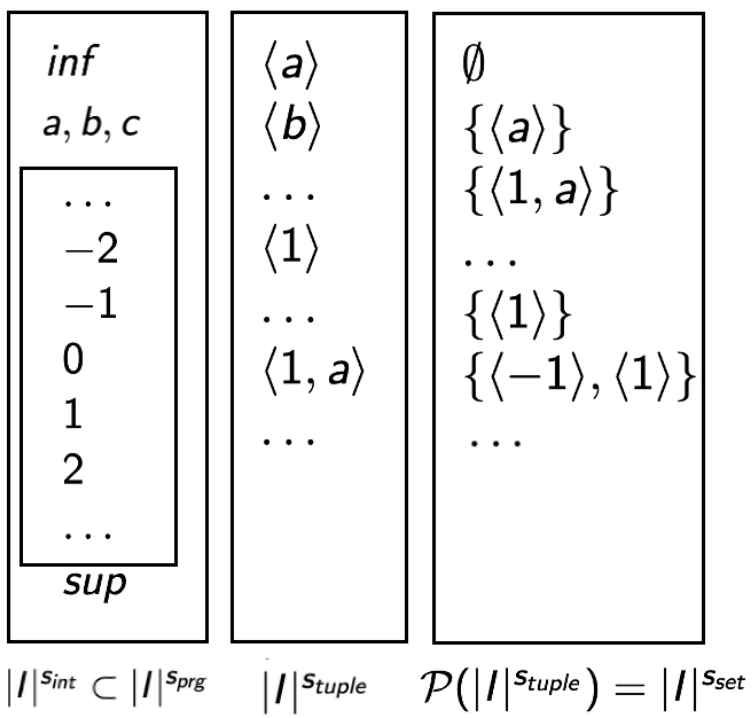}
    \caption{Universes corresponding to the sorts used in standard interpretations~\cite{fan22}.}
    \label{fig:universes}
\end{figure}

\paragraph{Sets and many-sorted logic in \anthem}
This is the last major task I hope to accomplish as part of my dissertation.
The question of integrating our aggregate semantics into \mg and \anthem is dependent upon our ability to (partially) axiomatize our notion of standard interpretations.
This special class of interpretations makes some assumptions about the behavior of the set sort (for instance, that set membership behaves in a standard way).
There is reason to believe this is possible -- \vampire natively supports a partial axiomatization of integer arithmetic~\cite{kovvor13a} which we extend to a partial axiomatization of two-sorted standard interpretations by including certain custom axioms in every verification task.
The Thousands of Problems for Theorem Provers (TPTP) project has numerous partial axiomatizations of theories compatible with \vampire, including some focusing on set theory~\cite{Sut17}.
We may be able to use these as a starting point, though the specifics (Figure~\ref{fig:universes}) of our many-sorted domain may be hard to express.
In  particular, we need to capture the restrictions that
\begin{enumerate}
    \item the numeral universe $|I|^{s_{int}}$ is a subsort of the program term universe $|I|^{s_{prg}}$,
    \item the tuple universe $|I|^{s_{tuple}}$ consists of tuples of program terms, and
    \item the set universe $|I|^{s_{set}}$ is the power set of the tuple universe.
\end{enumerate}
There is also the question of how theory extensions (such as set theory) impacts the runtime of \vampire and, consequently, the usability of \anthem.
Our experiments with \anthem and AP2P have already shown that certain programs containing integer arithmetic can be deceptively difficult to verify automatically.
For example, \anthem struggles to verify the external equivalence of 
$$
p(X*X) \ruleo ~X = 0..n.
$$
and 
$$
p(X*X) \ruleo ~X = -n..n.
$$
under the assumption that $n$ is an integer greater than $0$.

\section{Related Work}

The Introduction identifies three core challenges to ASP verification: modularity, grounding, and tool support.
Within each of these topics, there are a number of studies relevant to the research agenda proposed here.


\subsection{Modularity}

The notion of external equivalence presented earlier is closely related to modular equivalence for DLP and ELP-functions~\cite{janoiktomwol09}, which are in turn related to LP-functions~\cite[Section~2]{gel02}.
In these studies, the idea of program modules as composable functions mapping input atoms to output atoms (using hidden auxiliary atoms) has been explored in the context of disjunctive, propositional programs.
The \emph{module theorem}~\cite{janoiktomwol09} provides a compositional semantics for this class of programs.\footnote{Under certain reconfigurations of program modules, the module theorem can apply even when some input, output, or hidden atoms are forgotten~\cite{gonjanknoleiwol19}.}
Our work on external behavior of io-programs and modular arguments of correctness uses similar results established for a different class of programs, specifically, programs with variables and arithmetic.

\emph{Templates} are a construct roughly analogous to program modules, where global (public) predicates are renamed so as to interface with an enclosing program, and local (auxiliary/hidden/private) predicates are renamed with a procedure that (very nearly) guarantees the new identifiers are universally unique~\cite{alvianpaczan23}.
The use of templates promotes code reusability and eases the development of industrial-scale applications~\cite{calian06}.
This approach has the additional benefit of being able to test that simple invariants of templates hold in the context of an enclosing program.
%
As an alternative to the re-writing strategy described above, templates can also be defined as higher-order definitions of predicates~\cite{dashaljanden15}.
This is more in line with the second-order characterization of external behavior given by the $SM$ operator.

\subsection{Grounding}
Within the domain of translational semantics for ASP, a major distinction is between grounding-free approaches, and semantics applied to grounded or propositional programs.
A grounding-free translational approach is the basis of this proposal -- it benefits from being more general, but suffers some restrictions that have been overcome for propositional translations.
For example, the limitation of completion semantics to tight programs can be circumvented in the propositional case by adding loop formulas ($LF$) to the program's ($\Pi$) completion ($COMP(\Pi) \cup LF$)~\cite[Theorem 1]{linzhao04}.\footnote{The \textsc{assat} tool based on these results uses SAT solvers to compute stable models of propositional logic programs~\cite{linzhao04}.}
An extension of this theorem to the first-order case exists, but relies on a grounding procedure, namely, an instantiation of $COMP(\Pi) \cup LF$ w.r.t. a finite domain to obtain a propositional theory~\cite{chenlinwanzha06}.
Lee and Meng (2011) provide a full generalization of loop formulas to arbitrary first-order formulas -- this could be a promising extension to the completion procedure implemented by \anthem~\cite{leemeng11}.

\subsection{Tool Support}
Tools supporting formal verification of ASP programs can be roughly divided into the categories of \textbf{testing} based and \textbf{proof} based systems.
These are complementary approaches, since testing generally cannot provide the same level of assurance but is fast and universally applicable.
\textsc{harvey} is an ASP-based system for random testing of ASP programs~\cite{greoettom17}.
\aspide is an integrated development environment (IDE) supporting unit testing~\cite{ameberric21,febleorearic13}.
Similarly to \anthem, this supports modular verification by specifying conditions on outputs for program ``units" and automatically testing that these conditions are satisfied for certain inputs.
However, \anthem enables users to specify (possibly infinite) \emph{classes} of inputs rather than a finite set of test cases.


Examples of proof based systems are \selp~\cite{chenlinli05}, \tabeql~\cite{val04}, \textsc{lpeq}~\cite{janoik04}, and \textsc{ccT}~\cite{oetseitomwol09}.
\selp is closely related to this proposal, since it uses automated reasoning tools to check whether two logic programs are strongly equivalent.
One notable consequence of \selp's SAT checking methodology is its ability to find a counterexample, which would be an interesting addition to \anthem.
\selp differs from \anthem in its use of SAT solving instead of theorem proving; a more important difference is that \selp supports disjunctive propositional programs whereas \anthem supports normal programs with variables and arithmetic.
\textsc{tabeql} does not translate ASP programs into logical theories, but rather computes equilibrium models of arbitrary propositional theories using tableau calculi for here-and-there.
The ASP-to-ASP translation tool \textsc{LPEQ} and its variant \textsc{DLPEQ} produce logic programs whose answer sets (if any) represent counterexamples to the weak or strong equivalence of a pair of programs.\footnote{Two programs are \emph{weakly equivalent} if they have the same answer sets; this is a special case of external equivalence.}
Similarly to \selp, these systems accept disjunctive, variable-free logic programs.
A more flexible system is~\textsc{ccT}, which tests relativised strong equivalence with projection and uniform equivalence.\footnote{Uniform equivalence is a special case of strong equivalence where it is assumed that the context is a set of facts rather than an arbitrary set of rules.}
\textsc{ccT} is very similar in spirit to \anthem, \selp, and \textsc{LPEQ} given that it relies on a translation to quantified Boolean formulas evaluated by backend solvers.
This system is based on a general notion of program correspondence~\cite{eittomwol05}, and requires two sets of atoms (a \emph{context} and a \emph{projection set}) which behave similarly to \anthem's input and output predicates.
Again, \anthem is more general: rather than dealing with concrete sets of atoms, \anthem operates on classes of inputs defined by first-order assumptions.


\section{Conclusions and Future Directions}

\paragraph{ASP modules} I proposed this idea at the ICLP and LPNMR 2022 doctoral consortiums, and it received encouraging feedback from attendees.
Building off of the modular verification methodology discussed earlier, I would like to develop a repository of verified ASP sub-programs (``modules") that provide efficient, correct implementations of commonly encountered sub-problems.
Each module would be accompanied by a proof of correctness, whose guarantees can be used within an argument of the enclosing program's correctness.
This could reduce the effort needed for programming, and some of the labor required to formally prove the correctness of the program.
For example, defining the transitive closure of a binary relation is a common task in ASP programming.
Integrating a generic module such as
\begin{align*}
    transitive(X,Y) &\ruleo edge(X,Y).\\
    transitive(X,Z) &\ruleo transitive(X,Y), edge(Y,Z).
\end{align*}
is preferable to re-inventing the wheel with a custom, unverified implementation.
In this scenario, the programmer would only have to define the interfaces to and from the module, analogous to the input and output interfaces proposed for disjunctive logic programs~\cite{janoiktomwol09}.
Each program in the repository should have a natural language specification of intended behavior, a proof of correctness, and a description of how to interface with the module.
Thus, the idea is similar to that of ASP templates~\cite{alvianpaczan23}, with an emphasis on reusability of the corresponding proofs of correctness.
ASP practitioners could submit a (program, specification) pair as a verification challenge, or submit fully verified modules ready for reuse.
Besides the transitive closure module, other common, generalizable sub-problems include functional relations (predicates defining a mapping from one set to another) and definitions of grid adjacency (typically used in 2D planning problems like \textsc{asprilo}~\cite{geb18asprilo}).
While I'm still very interested in this task, I've come to realize that such a project is beyond the scope of my dissertation.
I now see it as a top priority for post-graduate research.
\paragraph{Discussion}
The task of verifying ASP programs is complex, but relating them to equivalent theories in the logic of here-and-there and many-sorted first-order logic is a promising approach. 
Much of my work has focused on investigating this relationship in the presence of advanced language constructs, such as aggregates and conditional literals. 
These are features on which modern ASP solutions rely heavily - yet they are not even the most sophisticated features offered by modern solvers.
Verification techniques for constraint answer set programs, or for programs with optimization statements and theory propagators, are possible future directions of research with great potential.

Another topic of interest is the extension of \anthem with alternative backend solvers.
Other ASP verification tools support finding counterexamples to program correspondence, which is an interesting and useful ability.
An SMT solver like CVC5~\cite{cvc522} might enable \anthem to check countersatisfiability and/or generate counterexamples.
Furthermore, intuitionistic theorem provers like nanoCOP-i~\cite{ott16} are a natural avenue to explore given ASP's close relationship with the logic of here-and-there.
Secondary transformations like completion could be avoided completely if an intuitionistic theorem prover -- possibly strengthened with axiom schemata such as~\eqref{eq:weak.ex.middle} -- shows itself capable of reasoning effectively within HTA.

Finally, there is the topic of making the verification task accessible enough that ASP practitioners will employ it in the real world.
Automation is a great way to promote this -- but much more work is required to make complex problems verifiable with reasonable resources.
Reusing components of programs and their corresponding proofs of correctness is another way to ease the burden on practitioners. 
To encourage formal verification in practice, \anthem could be integrated as a plugin to an IDE like \aspide in addition to behaving as a stand-alone tool.
I plan to continue working on such tool-assisted verification strategies in the future.

\bibliographystyle{eptcs}
\bibliography{bib}
\end{document}